\documentclass[10pt,canadian,english,journal=jpclcd,manuscript=letter,layout=twocolumn]{achemso}
\usepackage[T1]{fontenc}
\usepackage[latin9]{inputenc}
\usepackage{array}
\usepackage{multirow}
\usepackage{amsmath}
\usepackage{amssymb}
\usepackage{graphicx}

\makeatletter

\title{Interpretations of High-Order Transient Absorption Spectroscopies}

\author{Peter A. Rose}

\affiliation{Department of Physics, University of Ottawa, Ottawa, ON K1H 6N5,
Canada}

\author{Jacob J. Krich}

\affiliation{Department of Physics, University of Ottawa, Ottawa, ON K1H 6N5,
Canada}

\alsoaffiliation{School of Electrical Engineering and Computer Science, University
of Ottawa, Ottawa, ON K1H 6N5, Canada}

\email{jkrich@uottawa.ca}

\providecommand{\tabularnewline}{\\}

\usepackage{braket}
\usepackage{xr}

\makeatother

\usepackage{babel}
\begin{document}
\selectlanguage{canadian}%
\global\long\def\braket#1#2{\Braket{#1|#2}}%

\global\long\def\bra#1{\Bra{#1}}%

\global\long\def\ket#1{\Ket{#1}}%

\global\long\def\cvec#1{\boldsymbol{#1}}%

\begin{tocentry}
\includegraphics[width=5.1cm]{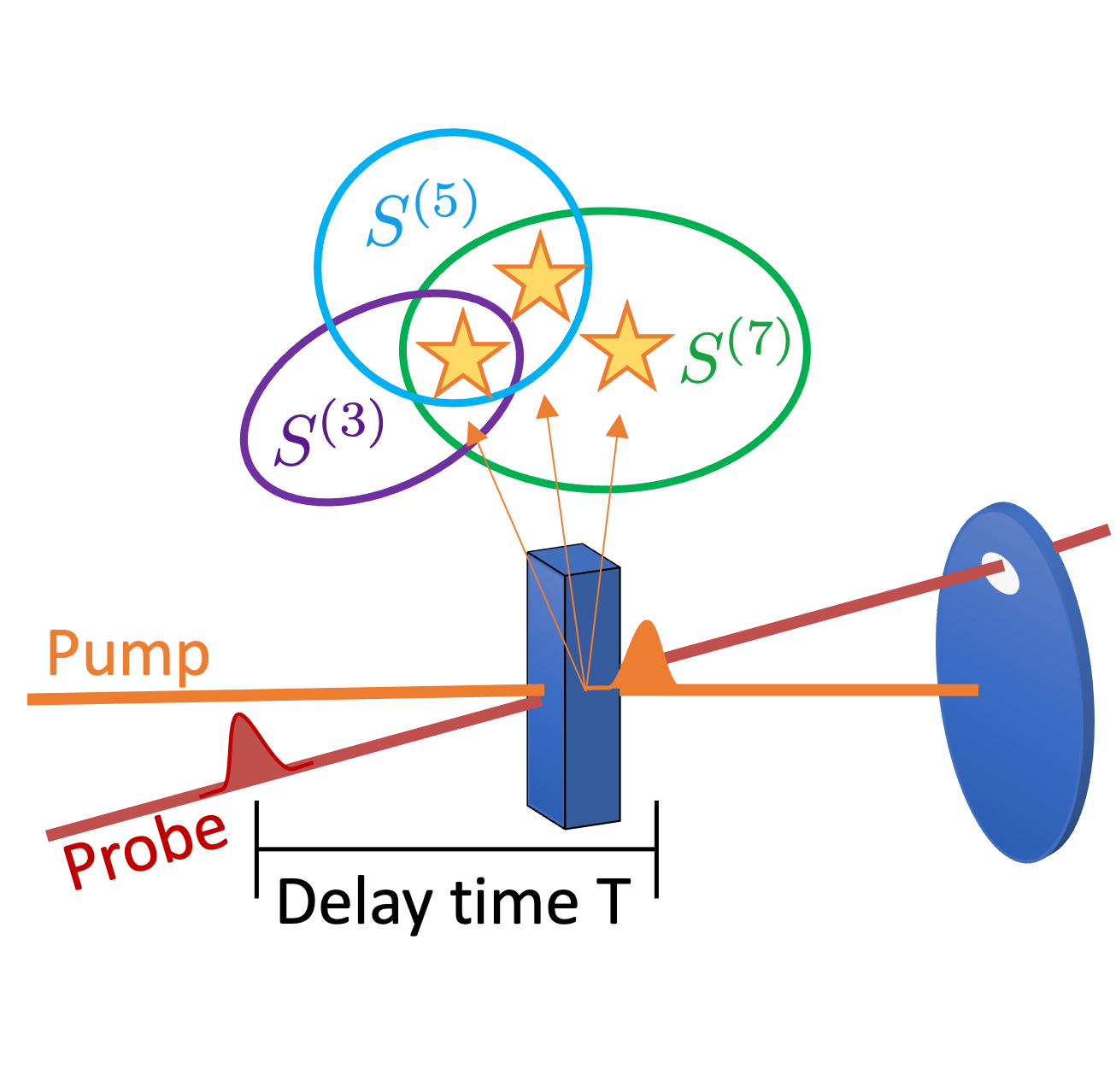}

\end{tocentry}
\begin{abstract}
Transient absorption (TA) spectroscopy has long been an invaluable
tool for determining the energetics and dynamics of excited states
in atomic, molecular, and solid-state systems. When pump pulse intensities
are sufficiently high, the resulting TA spectra include both the generally
desired third-order response of the studied material as well as responses
that are higher order in the electric field amplitudes of the pulses.
It has recently been shown that pump-intensity-dependent TA measurements
allow separating the various orders of response of the TA signal,
but the information content available in those higher orders has not
been described. We give a general framework, intuition, and nomenclature
for understanding the information contained in high-order TA spectra.
Standard TA spectra are generally interpreted in terms of three fundamental
processes: ground-state bleach (GSB), stimulated emission (SE), and
excited state absorption (ESA), and we extend those concepts to higher
order. Each order introduces two new processes: SE and ESA from highly
excited states that were not accessible in lower orders. In addition,
each order contain negations of lower-order processes, just as GSB
is a negation of the linear absorption. We show the new spectral and
dynamical information that is introduced at each order and show how
the relative signs of the signals in different orders can be used
to identify which processes are dominant.
\end{abstract}
\selectlanguage{english}%

\mciteErrorOnUnknownfalse 

New forms of spectroscopy give new insights into the properties and
dynamics of systems from atoms to solids. Linear absorption spectroscopy,
in frequency ranges from THz to x-ray, gives information about the
ground or thermal states of a material and its optically accessible
excited states. Transient absorption (TA) spectroscopy measures the
change in a probe pulse's absorption spectrum when it arrives a time
$T$ after a pump pulse, revealing both the absorption and dynamics
of excited states \cite{jonas1995,berera2009,hamm2009,ohkita2011,cho2013,maiuri2019,pandya2020,provazza2021,cina2022,xu2022,unger2022,sayer2023}.
The recently developed high-order transient absorption (HOTA) spectroscopy
extends TA spectroscopy by systematically separating higher orders
of nonlinear response, which has not previously been possible using
TA methods \cite{maly2023,luttig2023,luttig2023a}. These higher orders
of nonlinear response contain both spectral and dynamical information
about multiply excited states.

TA signals with pump pulses that are not too strong can be intuitively
understood in terms of three fundamental processes: excited state
absorption (ESA), stimulated emission (SE), and ground state bleach
(GSB) \cite{mukamel1999}. There is a small probability that the pump
excites the system into any of the available single-excitation states,
which then evolve in time. When the probe arrives, the singly excited
populations can either undergo SE back to the ground state or ESA
up to doubly excited states. The population in the excited state is
associated with a reduced population in the ground state, producing
the GSB, which is a reduction of the linear absorption from the level
without the pump.

The TA signal $S(T,\omega)$ gives the change in absorption of a probe
pulse at frequency $\omega$, which depends on the intensity of the
pulses. Taking the electric dipole approximation and treating the
fields classically, the standard interaction Hamiltonian is $H^{\prime}(t)=-\mu\cdot\mathbf{E}(t)$,
where $\mathbf{E}(t)=\text{\ensuremath{\mathbf{E}_{a}(t)}}+\mathbf{E}_{b}(t)$
is the electric field of the pump ($a)$ and probe ($b$) pulses.
We write $\cvec E_{j}=\lambda_{j}[\cvec e_{j}\varepsilon_{j}(t)+\cvec e_{j}^{*}\varepsilon_{j}^{*}(t)]$,
with $\cvec e_{j}$ the polarization and $\varepsilon_{j}(t)$ containing
the pulse amplitude, wavevector, and absolute phase. For our purposes,
the key feature is the dimensionless $\lambda_{j}$, which scales
the amplitude of the pulse. The typical third-order signal derives
from a polarization field that is proportional to $\lambda_{a}^{2}\lambda_{b}$.
We define a dimensionless pump intensity $I=\lambda_{a}^{2}$.

Moving beyond the lowest-order response, $S$ can be expressed as
a power series in $I$, 
\begin{align}
S(T,\omega,I) & =\overbrace{S^{(3)}(T,\omega)I}^{\text{standard TA signal}}+S^{(5)}(T,\omega)I^{2}\nonumber \\
 & +S^{(7)}(T,\omega)I^{3}+....\label{eq:perturbative-expansion}
\end{align}
\textbf{ }Eq.~\ref{eq:perturbative-expansion} assumes that the probe
is weak and unchanging, and $S^{(n)}$ gives the $n^{\text{th}}$
order material response, with only odd orders contributing to TA of
samples large enough for phase matching \cite{mukamel1999,boyd2008}.\textbf{
}Standard TA either chooses $I$ small enough that the contributions
beyond $S^{(3)}$ are negligible or has the higher-order contributions
uncontrollably mixed with the lower-order ones, allowing study of
$I$-dependent trends \cite{auston1975,pedersen1993,bittner1994,smith1994,valkunas1995,yokoyama1998,klimov2000,bruggemann2004,ueda2008,hoffmann2009,taguchi2011,almand-hunter2014,chlouba2019,soni2021,navotnaya2022,kumar2023}.
The new HOTA spectroscopy allows separating the $S^{(n)}$ at an intensity
$I_{0}$ by combining TA spectra taken with $N$ pump intensities
\[
I_{p}=4I_{0}\cos^{2}\left(\frac{\pi p}{2N}\right)
\]
for $p=0\dots N-1$, where $I_{0}$ is chosen so that the contribution
from $S^{(2N+3)}$ is negligible. The full method is described in
Refs.~\citenum{maly2023} and \citenum{luttig2023}.

The newly accessible HOTA spectroscopies contain information that
has not previously been accessible in TA. This manuscript extends
the standard ESA/SE/GSB pathways to higher orders, showing the spectral
and dynamical information available in each new order in HOTA. We
begin by discussing the dynamical information, which is easier to
understand and serves as an intuitive starting place for discussing
the higher-order signals. We then describe all of the Liouville pathways
that exist for each order, introducing a nomenclature to discuss those
contributions that is as useful as ESA/SE/GSB. We then describe the
spectral information that exists in each order, focusing on two illustrative
model systems.

\begin{figure}
\includegraphics{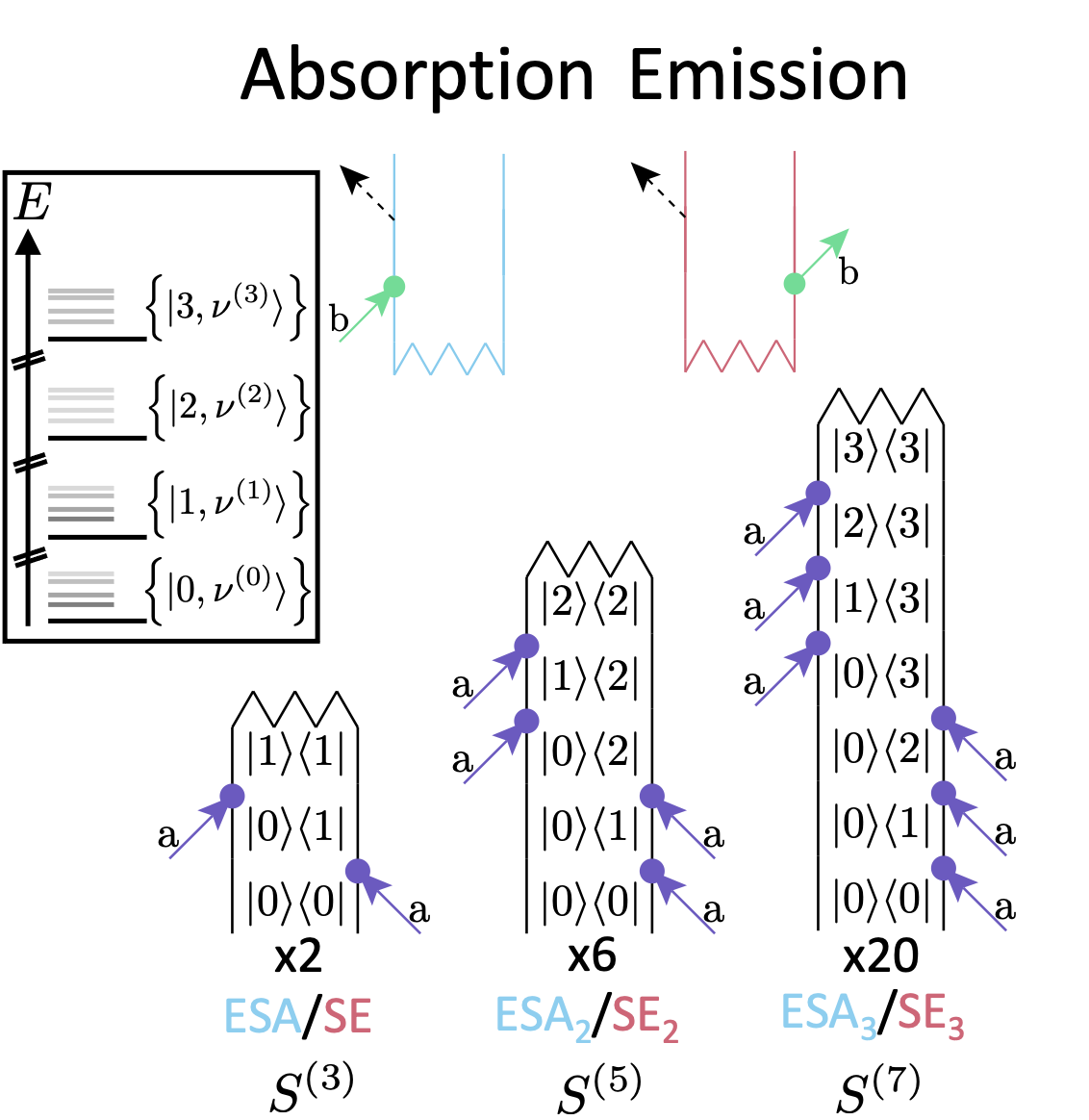}\caption{\label{fig:SEnESAn} Sample Feynman diagrams for high-order processes
broken into \emph{bases} showing the action of the pump pulse, labeled
``$a$'', and \emph{caps} showing the action of the probe pulse,
labeled ``$b$''. Each base pairs with each cap, resulting in a
complete Feynman diagram. Bases shown produce the highest possible
number of excitations at third-, fifth-, and seventh-order, producing
the ESA$_{n}$/SE$_{n}$ diagrams with the appropriate cap. The multiplicative
number under each base indicates the number of similar bases formed
by permutation of the ordering of the pump arrows, confirmed with
an automated diagram generator \cite{rose2021a}. The excitation numbers
are not shown in the caps, since the caps pair with every base and
because $n$ can change due to relaxation during the delay time $T$.
Inset: Sample ladder-type spectrum, with 0-, 1-, 2-, and 3-excitation
states shown.}
\end{figure}

We make a standard set of assumptions about the systems and light
pulses under study. Before the pump arrives, the system is in a thermal
state, $\rho_{\text{th}}$. We describe systems -- as are commonly
discussed in optical spectroscopy -- with a ladder of dipole-allowed
transitions, where $n$-times excited states are distinct from $(n+1)$-times
excited states, as indicated in the inset of Fig.~\ref{fig:SEnESAn}.
We use $\ket n$ to refer to the set of $n$-times excited states
and use $\ket{n,\nu^{(n)}}$ to refer to a particular state. Such
an excitation ladder requires that the states $\left\{ \ket{n,\nu^{(n)}}\right\} $
are long-lived with respect to the pump pulse duration. If this separation
of time scales does not occur, Eq.~\ref{eq:perturbative-expansion}
is still valid, but much of the intuition developed in this paper
is not directly applicable.

Each signal $S^{(2n+1)}(T)$ in Eq.~\ref{eq:perturbative-expansion}
can contain the dynamics of at most $n$-times excited states. This
intuitive claim follows from the excitation ladder model and the TA
phase-matching condition \cite{mukamel1999,hamm2011}.\textbf{ }We
illustrate it using double-sided Feynman diagrams to describe the
pathways that contribute to $S^{(2n+1)}$. We assume that the probe
is weak, so there is only one interaction with the probe, the remaining
$2n$ interactions are with the pump, and the two pulses do not overlap
in time. TA phase matching requires that there be equal numbers of
left- and right-directed pump interactions, representing the $\varepsilon_{a}$
and $\varepsilon_{a}^{*}$ terms in the field, and the rotating wave
approximation (RWA) requires that arrows directed in cause excitations
while arrows directed out cause de-excitations \cite{mukamel1999,hamm2011}.
Then the highest excitation states that can be reached after the pump
occur when the raising operator is applied $n$ times to both the
bra- and ket-sides of the initial density matrix, which requires $2n$
interactions. We divide the Feynman diagrams into \emph{bases} that
involve only interactions with the pump pulse and two\emph{ caps}
that involve only one interaction with the probe, as shown in Fig.~\ref{fig:SEnESAn},
with a full diagram formed by combining any base with either cap.\textbf{
}The bases shown in Fig.~\ref{fig:SEnESAn} represent the highest-excitation
pathways available at third, fifth, and seventh order. Lower excitation
pathways also exist, and we discuss them below. In Fig.~\ref{fig:SEnESAn},
the label $\ket n\bra m$ is a placeholder for a density matrix $\sum_{jk}c_{jk}\ket{n,\nu_{j}^{(n)}}\bra{m,\nu_{k}^{(m)}}$.

We generalize the concept of stimulated emission and excited state
absorption from the singly excited states to similar processes that
originate from the $n$-excitation states. When the $S^{(3)}$ bases
shown in Fig.~\ref{fig:SEnESAn} are combined with the absorption
(emission) cap we obtain the familiar ESA (SE) pathway. By analogy,
when the $S^{(5)}$ bases are combined with the absorption (emission)
cap, we obtain a conceptually similar type of signal, which is excited-state
absorption or stimulated emission from the doubly excited states.
We call these pathways $\text{ESA}_{2}/\text{SE}_{2}$ for $S^{(5)}$
and $\text{ESA}_{n}/\text{SE}_{n}$ for higher orders. These two pathways
contain all of the new dynamical information that is revealed at order
$(2n+1)$. When $T$ is greater than the lifetime of the $n$-times
excited states, the $n$-times excited population decays during the
waiting time, and the $\text{ESA}_{n}\left(\text{SE}_{n}\right)$
pathways include absorption (emission) processes beginning from lower-excitation
states. Such decays generally change the signal, allowing those lifetimes
to be determined \cite{maly2023}. Each HOTA signal also involves
contributions from pathways with lower numbers of excitations immediately
after the pump. Those pathways have dynamical information that generally
already exists in lower-order signals, though interferences of the
type discussed below could mask the signal in lower orders. The exception
to this rule occurs at third order, where GSB, which is a negation
of linear absorption, also introduces ground-state vibrational dynamics
that were not present in the static linear absorption signal. As usual,
care must be taken in assigning physical significance to pathways.
For example, in a two-level system with relaxation, there is no physical
excited-state absorption process. However, the ESA pathway must still
be calculated, because after the pump pulse excites the system, the
system relaxes to the ground state. The probe pulse can then excite
the system again. The dynamical information about the lifetime of
the excited state is still captured in the $\text{SE}/\text{ESA}$
pathways.

Since each order controllably adds a single set of excitation states
not in the lower-order signals, new dynamics in higher orders can
immediately be attributed to new processes. One can analyze each order
sequentially, beginning from third order and characterizing the decay
constants (e.g., using global analysis \cite{knutson1983,steinbach2002,vanstokkum2004})
and beat frequencies associated with each order. If the fifth-order
signal contains the same decay constants and frequencies as the third-order
and also contains a new set of dynamics, those new dynamics can be
attributed to processes involving only the states $\ket{2,\nu^{(2)}}$,
and analogously for higher orders.

We now consider the full set of pathways contributing to TA signals
at all orders. We build up an understanding by comparing the well-known
third-order signal to the linear absorption signal (A).\textbf{ }Standard
third-order TA involves two new processes involving singly excited
states after the pump (ESA/SE) and a negation of the linear absorption
(GSB). In order to generalize to higher orders, we also refer to GSB
as negated absorption (NA$^{(3)}$). This labeling introduces a pattern
that applies at higher orders: the ($2n+1$)-order signal introduces
two new processes (ESA$_{n}$/SE$_{n}$) and negations of all lower-order
processes, which we illustrate in Table~\ref{tab:all-pathways} and
now describe.

\begin{figure}
\includegraphics{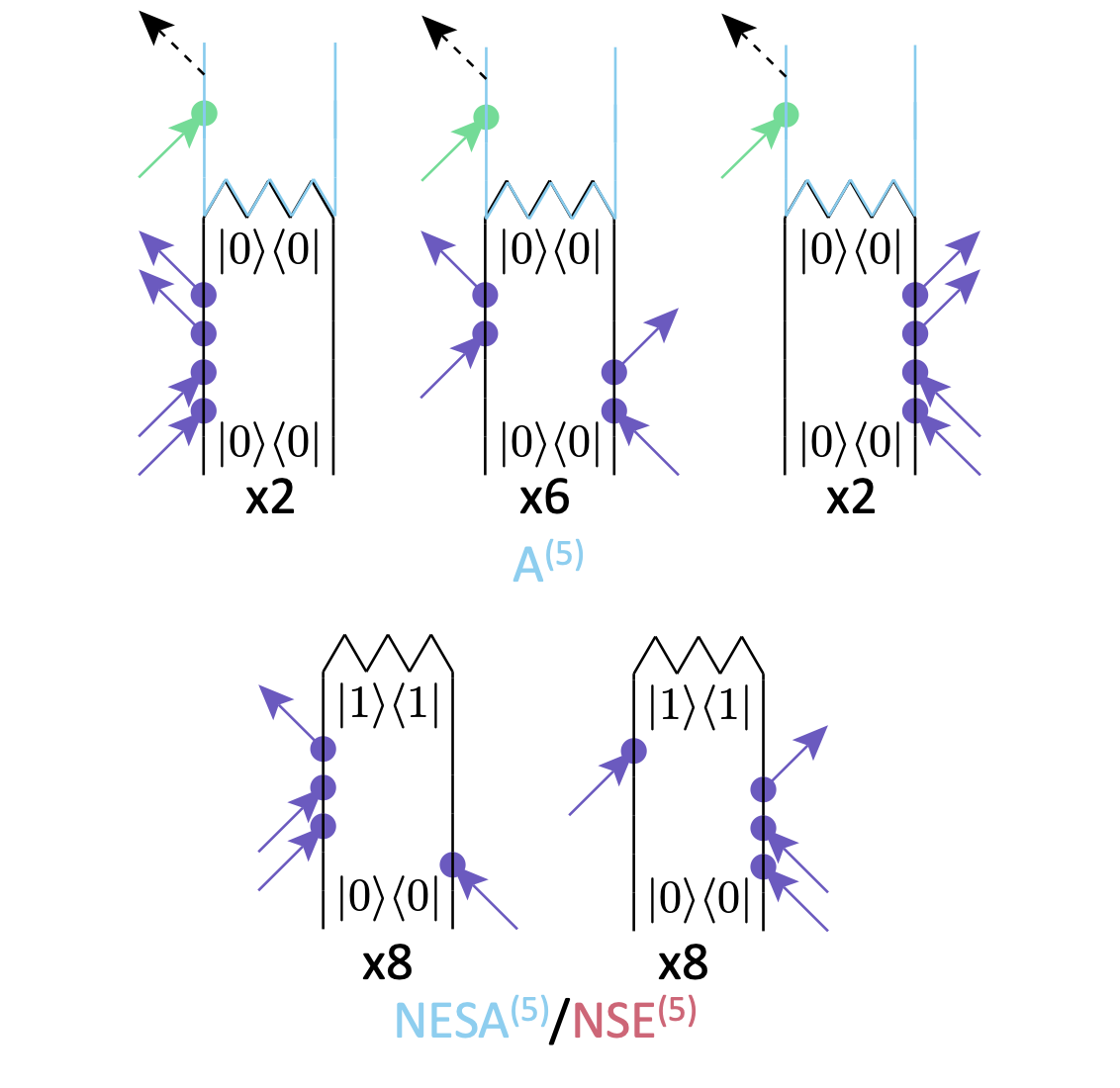}\caption{\label{fig:5th-order} Pathways contributing to $S^{(5)}$ that negate
third-order processes. Top row shows contributions to increased ground-state
absorption and include the absorption cap. Second row shows bases
associated with the negation of the third-order ESA and SE pathways.
All 16 bases contribute to $\text{NESA}^{(5)}$ ($\text{NSE}^{(5)}$)
when paired with the absorption (emission) cap shown in Fig.~\ref{fig:SEnESAn}.
The number of similar permutations of the pump interactions is below
each diagram. The diagrams associated with $\text{ESA}_{2}/\text{SE}_{2}$
are displayed in Fig.~\ref{fig:SEnESAn}.}
\end{figure}

Next we compare the fifth- and third-order signals. The fifth-order
signal contains new information about the 2-excitation states in the
$\text{ESA}_{2}/\text{SE}_{2}$ pathways shown in Fig.~\ref{fig:SEnESAn}.
It also includes processes that can be interpreted as negations of
the signals associated with the 1-excitation states, which we call
negated excited state absorption ($\text{NESA}$) and negated stimulated
emission ($\text{NSE}$), shown diagrammatically in Fig.~\ref{fig:5th-order}.
Intuitively these negations occur because the fifth-order response
involves moving population from the 1-excitation states that produce
ESA/SE to the 2-excitation states as well as back to the 0-excitation
states. That increase in ground state population gives an increase
in the linear-absorption-type process (undoing part of the GSB), which
we call $\text{A}^{(5)}$ .

\begin{figure}
\includegraphics{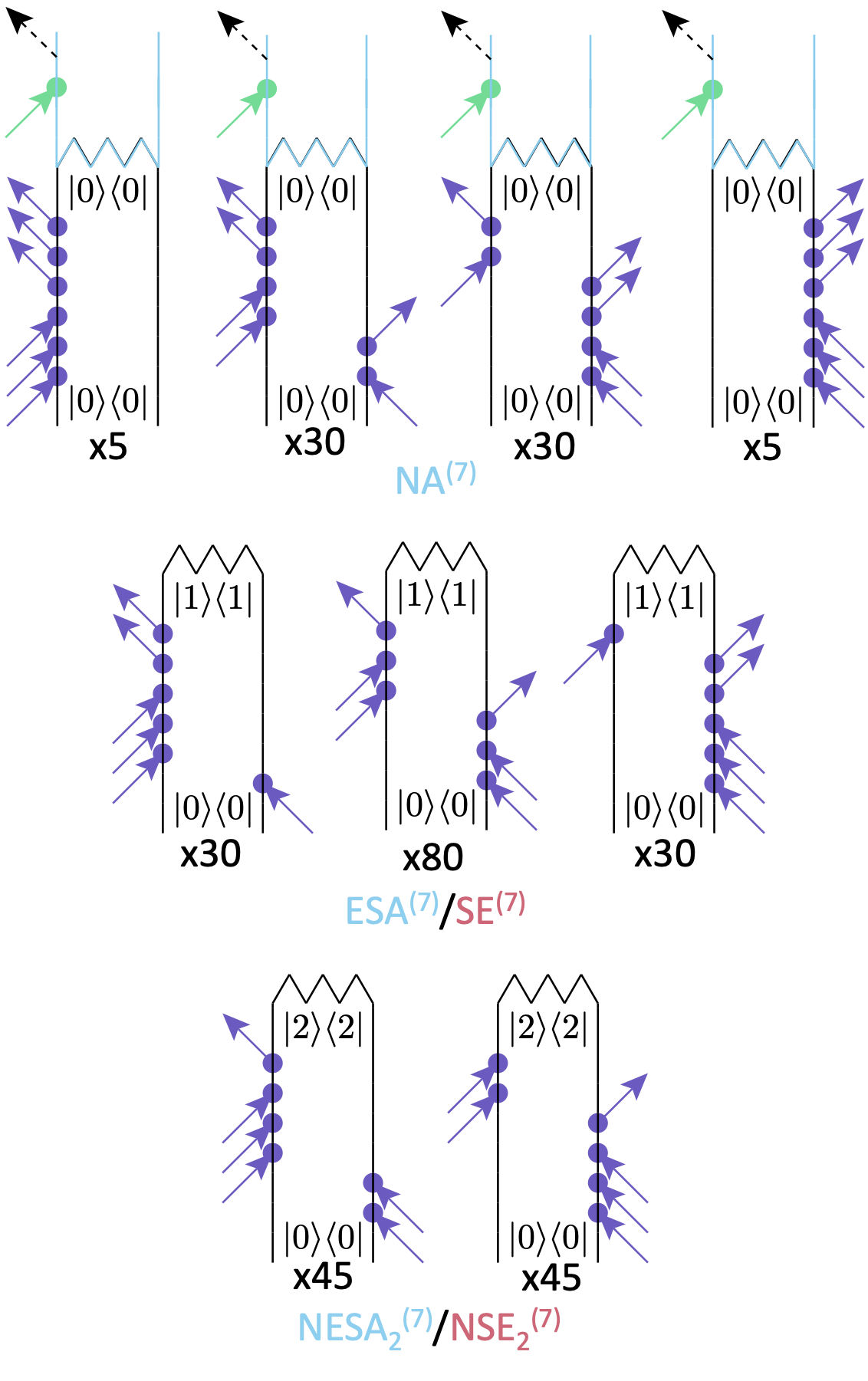}\caption{\label{fig:7th-order}Pathways contributing to $S^{(7)}$ that negate
fifth-order processes. Top row shows contributions to negated ground-state
absorption and include the absorption cap. Second row shows bases
associated with the increase of the third-order ESA and SE processes.
Third row shows bases associated with the negation of the fifth-order
$\text{ESA}_{2}$ and $\text{SE}_{2}$ processes. The number of similar
permutations of the pump interactions is below each diagram. The bases
associated with $\text{ESA}_{3}/\text{SE}_{3}$ are displayed in Fig.~\ref{fig:SEnESAn}.}
\end{figure}

We can extend this pattern and nomenclature to any order, which we
demonstrate by comparing the seventh- and fifth-order signals. The
new information in seventh order is contained in the $\text{ESA}_{3}/\text{SE}_{3}$
processes, which report on the 3-excitation states. The seventh-order
signal also includes negations of those in the fifth-order pathways,
with the relevant bases shown in Fig.~\ref{fig:7th-order}. The seventh-order
signal includes negations of the $\text{ESA}_{2}/\text{SE}_{2}$ pathways,
which we call $\text{NESA}_{2}^{(7)}/\text{NSE}_{2}^{(7)}$, a contribution
to the $\text{ESA}/\text{SE}$-type processes, which have the same
signs as the originals and which we call $\text{ESA}^{(7)}/\text{SE}^{(7)}$,
and a negation of the absorption process, which we call $\text{NA}^{(7)}$.

Each higher-order signal contains pathways that are related to the
pathways at the previous order but have opposite signs, as shown on
the sub-diagonal of Table~\ref{tab:all-pathways}\textbf{. }We gave
the intuitive explanation for these alternating signs; diagrammatically,
the alternation can be understood by analyzing the bases in Figs.~\ref{fig:SEnESAn},
\ref{fig:5th-order} and \ref{fig:7th-order}. Each signal $S^{(2m+1)}$
is proportional to a factor of $(-i)^{2m+2}$ , so every signal order
$S^{(2m+1)}$ has an overall factor of $i^{2}=-1$ as compared with
$S^{(2m-1)}$.\cite{mukamel1999} In addition, every arrow that occurs
on the bra side of the diagrams gives a factor of $-1$. Combining
these two rules gives the conclusion that $\text{ESA}_{n}$ has the
same sign for all $n$, and similarly that $\text{SE}_{n}$ has the
same sign for all $n$. However, all of the diagrams that correspond
to signals that existed in the lower order involve adding an even
number of arrows to the bra side of the diagram, as compared to the
previous order. Therefore all of these diagrams contribute the opposite
sign relative to their previous-order counterparts. The supplementary
information contains a discussion of modifications that are required
when pump or probe photon energies are close to the thermal energy,
e.g., with THz spectroscopies, as well as a discussion of the permutations
of the diagrams shown in Figs~\ref{fig:SEnESAn}-\ref{fig:7th-order}.
\begin{table}
\caption{\label{tab:all-pathways} Names of pathways in (transient) absorption
spectroscopy to seventh order. The $\pm$ shows the sign of the pathway's
contribution. Pathways representing new dynamical information are
in bold and appear along the diagonal. Pathways left of the diagonal
negate (with the letter N) or revive pathways that first appeared
at lower orders. Superscript shows the order of the contribution,
which is suppressed for the diagonal contributions }

\footnotesize

\begin{tabular}{c|llll}
\multicolumn{1}{c}{} &  &  &  & \tabularnewline
\hline 
\hline 
\multicolumn{1}{c}{} & \multicolumn{4}{c}{Optical excitations after pump}\tabularnewline
\multicolumn{1}{c}{Order} & \multicolumn{1}{c}{0} & \multicolumn{1}{c}{1} & \multicolumn{1}{c}{2} & \multicolumn{1}{c}{3}\tabularnewline
\cline{2-5} \cline{3-5} \cline{4-5} \cline{5-5} 
\multirow{2}{*}{1} & \multirow{2}{*}{\textbf{$+\ensuremath{\phantom{N}}\mathbf{\text{A}}$}} &  &  & \tabularnewline
 &  &  &  & \tabularnewline
\multirow{2}{*}{3} & $-\text{NA}^{(3)}$ & $+\ensuremath{\phantom{N}}\mathbf{\text{\textbf{ESA}}}$ &  & \tabularnewline
 & $\phantom{-}\text{(GSB)}$ & $-\ensuremath{\phantom{N}}\mathbf{\text{\textbf{SE}}}$ &  & \tabularnewline
\multirow{2}{*}{5} & \multirow{2}{*}{$+\ensuremath{\phantom{N}}\text{A}^{(5)}$} & $-\text{NESA}^{(5)}$ & $+\ensuremath{\phantom{N}}\mathbf{\mathbf{\text{\textbf{ESA}}}_{\mathbf{2}}}$ & \tabularnewline
 &  & $+\text{NSE}^{(5)}$ & $-\ensuremath{\phantom{N}}\mathbf{\text{\textbf{SE}}_{2}}$ & \tabularnewline
\multirow{2}{*}{7} & \multirow{2}{*}{$-\text{NA}^{(7)}$} & $+\text{\ensuremath{\phantom{N}}ESA}^{(7)}$ & $-\text{NESA}_{2}^{(7)}$ & $+\ensuremath{\phantom{N}}\mathbf{\text{\textbf{ESA}}_{3}}$\tabularnewline
 &  & $-\text{\ensuremath{\phantom{N}}SE}^{(7)}$ & $+\text{NSE}_{2}^{(7)}$ & $-\ensuremath{\phantom{N}}\mathbf{\text{\textbf{SE}}_{3}}$\tabularnewline
\hline 
\hline 
\multicolumn{1}{c}{} &  &  &  & \tabularnewline
\end{tabular}
\end{table}

To demonstrate some of the new spectral information present in HOTA,
we introduce two model systems. The first is a ladder model with Hamiltonian
\begin{equation}
H_{0}=\hbar\left(\begin{array}{ccccc}
0 & 0 & 0 & 0 & 0\\
0 & \omega_{1} & 0 & 0 & 0\\
0 & 0 & \omega_{2} & 0 & 0\\
0 & 0 & 0 & \omega_{3} & 0\\
0 & 0 & 0 & 0 & \omega_{4}
\end{array}\right),\label{eq:H_ladder}
\end{equation}
and dipole operator 
\begin{equation}
\mu=\left(\begin{array}{ccccc}
0 & \mu_{10} & 0 & 0 & 0\\
\mu_{10} & 0 & \mu_{21} & 0 & 0\\
0 & \mu_{21} & 0 & \mu_{32} & 0\\
0 & 0 & \mu_{32} & 0 & \mu_{43}\\
0 & 0 & 0 & \mu_{43} & 0
\end{array}\right).\label{eq:mu_ladder}
\end{equation}
We take $\mu_{n,n-k}$ for $k>1$ to be zero because we assume that
the pulse bandwidth does not support such transitions. The frequency
differences $\omega_{n,n-1}\equiv\omega_{n}-\omega_{n-1}$ are the
spectral locations where peaks may appear. Note that this excitation
ladder model contains no dynamics, so the signal $S^{(2n+1)}$ is
constant for $T>0$, in the impulsive limit. This model could represent
an anharmonic vibrational mode of a molecule, which might be studied
using IR pump and probe pulses \cite{hamm2011}. The purpose of this
discussion is to paint a general picture of the kinds of spectral
information that can be revealed (or hidden) in HOTA spectra, regardless
of the wavelengths or the sample type.

\begin{figure}
\includegraphics{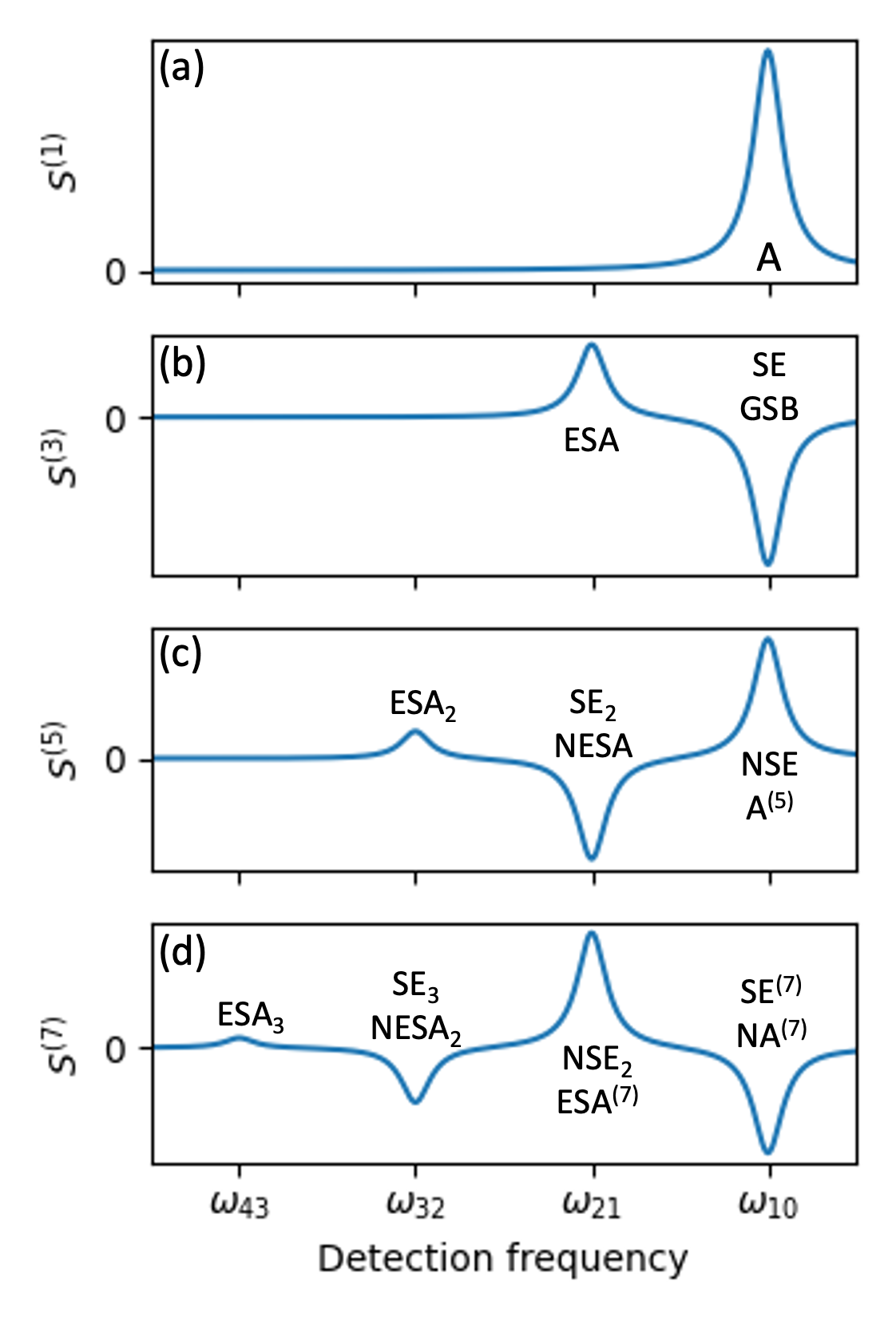}\caption{\label{fig:AnharmonicSpectra}Simulated spectra of an anharmonic ladder
for (a) linear absorption (b) standard (third-order) TA, (c) fifth-order
TA and (d) seventh-order TA. The $\text{ESA}_{n}$ process is the
lowest-energy peak in each spectrum, and always contributes with a
positive sign. All spectral features from one spectrum appear with
a negative sign in the next spectrum.}
\end{figure}

At each order, new spectral information is revealed by the $\text{ESA}_{n}$
process, which reports on energetic transitions that were inaccessible
at the previous order. As long as the new energetic transitions are
distinct from previous transitions, this new information appears in
spectra as new, positive-signed peaks. To illustrate this idea, we
show simulated spectra for an unequally spaced ladder model in Fig.~\ref{fig:AnharmonicSpectra},
for which we use $\{\omega_{1},\omega_{2},\omega_{3},\omega_{4}\}=\{1,1.9,2.7,3.4\}\omega_{0}$
and have assumed that the excitation energies are larger than the
thermal energy, with simulations using the open-source ultrafast spectroscopy
suite (UFSS) \cite{rose2019,rose2021,rose2021a}. We take all $\mu_{n,n-1}=1$
and simulate a closed system with no relaxations, with all of the
peaks given the same phenomenological linewidths for visualizing the
spectra. The peak at $\omega_{10}$ corresponds to linear absorption
at first-order, GSB/SE at third order, and the negations/revivals
of these processes at higher orders. The sign of the peak at $\omega_{10}$
flips with each order. In addition, each order reveals a new spectral
peak, corresponding to ESA at third order and $\text{ESA}_{n}$ at
higher-order. Peaks at the same spectral positions appear at higher
orders, again with alternating sign.

When spectral peaks overlap, interpreting the spectra becomes more
complicated. To build intuition, we move to a degenerate ladder model
in which the ESA$_{n}$/SE$_{n}$ processes are all at the same frequency
and allow $\mu$ in Eq.~\ref{eq:mu_ladder} to vary. In the supplementary
information we derive expressions for the third- and fifth-order signals,
using a model with no 3-excitation states ($\mu_{32}=0$) and with
$\omega_{10}=\omega_{21}=\omega_{0}$. We see that 
\begin{align}
S^{(3)}(T,\omega_{0}) & \propto-2\mu_{10}^{4}+\mu_{21}^{2}\mu_{10}^{2},\label{eq:S^3}\\
S^{(5)}(T,\omega_{0}) & \propto-\frac{1}{12}\left(7\mu_{21}^{4}\mu_{10}^{2}-\mu_{21}^{2}\mu_{10}^{4}-8\mu_{10}^{6}\right).\label{eq:S^5}
\end{align}
In this case, the third-order signal vanishes when $\mu_{21}=\sqrt{2}\mu_{10}$,
whereas the fifth-order signal vanishes when $\mu_{21}=\sqrt{8/7}\mu_{10}$,
assuming that all processes give the same lineshapes. In the range
$\sqrt{8/7}<\mu_{21}<\sqrt{2}$, both the third- and fifth-order signals
are negative, breaking the expected pattern of alternating signs that
appears in Fig.~\ref{fig:AnharmonicSpectra}. Note that, for a harmonic
vibrational mode, $\mu_{21}=\sqrt{2}\mu_{10}$, and the third-order
peak vanishes \cite{hamm2009}. A third-order peak that vanishes due
to such interferences between pathways is not unique to vibrational
modes, and we demonstrate below how a simple biexciton model may produce
the same effect.

If a negative-signed peak appears in both the third- and fifth-order
signals at short $T$, it indicates overlap of transitions to 1- and
2-excitation states. Negative peaks in third-order signals originate
in GSB/SE processes between the 0- and 1-excitation states. Looking
at Table \ref{tab:all-pathways}, the only negative contributions
to $S^{(5)}$ are SE$_{2}$ and NESA$^{(5)}$, which both occur at
the same spectral positions: transitions between 1- and 2-excitation
states. The fifth-order negations of the GSB/SE processes have positive
sign, so if the fifth-order signal at the same spectral position is
also negative, it indicates the presence of a hidden ESA process at
third order and indicates that the dipole contributing to that ESA
pathway, $\mu_{21}$, cannot be too large. In this case, the fifth-order
spectra gives new spectral information about 1-excitation states,
as we found in the study of quantum dots in Ref.~\citenum{maly2023}.

As an interesting limiting case, if a peak at some frequency first
appears in $S^{(5)}$ and it has a negative amplitude, it reveals
a dipole-allowed transition between both $\ket 0\rightarrow\ket 1$
and $\ket 1\rightarrow\ket 2$ that are normally visible in third
order but canceled each other. This situation is in contrast to a
peak that first appears in $S^{(5)}$ with positive sign, which reports
on transitions between $\ket 2$ and $\ket 3$ states, as discussed
above. In our simple model, a new negative peak demonstrates that
$\mu_{21}$ must be roughly $\sqrt{2}$ stronger than $\mu_{10}$.

This phenomenon of 1-excitation pathways canceling in the third-order
signal while appearing in the fifth-order signal can occur in many
situations, and we now introduce a model of optical spectroscopy of
biexcitons to illustrate another simple case. Let $a_{A}^{\dagger}$
and $a_{B}^{\dagger}$ create excitons and let the Hamiltonian be
\[
H_{0}=\hbar\sum_{i=A,B}a_{i}^{\dagger}a_{i}\omega_{i}+J\left(a_{A}^{\dagger}a_{B}+h.c.\right),
\]
where $\hbar$ is Planck's constant and $J$ is the coupling between
the two bare excitons with energies $\hbar\omega_{i}$. The ground-state
energy is $\hbar\omega_{g}=0$, and the biexciton has energy $\hbar\omega_{AB}=\hbar\left(\omega_{A}+\omega_{B}\right)$.
We set the dipole transitions for each bare exciton equal, that is
$\mu_{g,A}=\mu_{g,B}\equiv\mu_{0}$, and similarly $\text{\ensuremath{\mu_{A,AB}=\mu_{B,AB}}=}\mu_{0}$
, which is different from the previous ladder model, in which the
third-order signal vanished only if $\mu_{21}=\sqrt{2}\mu_{10}$.
Let $\hbar\omega_{\alpha}$ and $\hbar\omega_{\beta}$ be the eigenenergies
of the single exciton states, with corresponding transition dipoles
$\mu_{g,\alpha},\mu_{g,\beta},\mu_{\alpha,AB}$ and $\mu_{\beta,AB}$.
Tuning $J$ modifies the single exciton eigenstates and, crucially,
$\mu_{g,\alpha},\mu_{g,\beta},\mu_{\alpha,AB}$ and $\mu_{\beta,AB}$
\cite{yuen-zhou2014}.

\begin{figure}
\includegraphics[width=1\columnwidth]{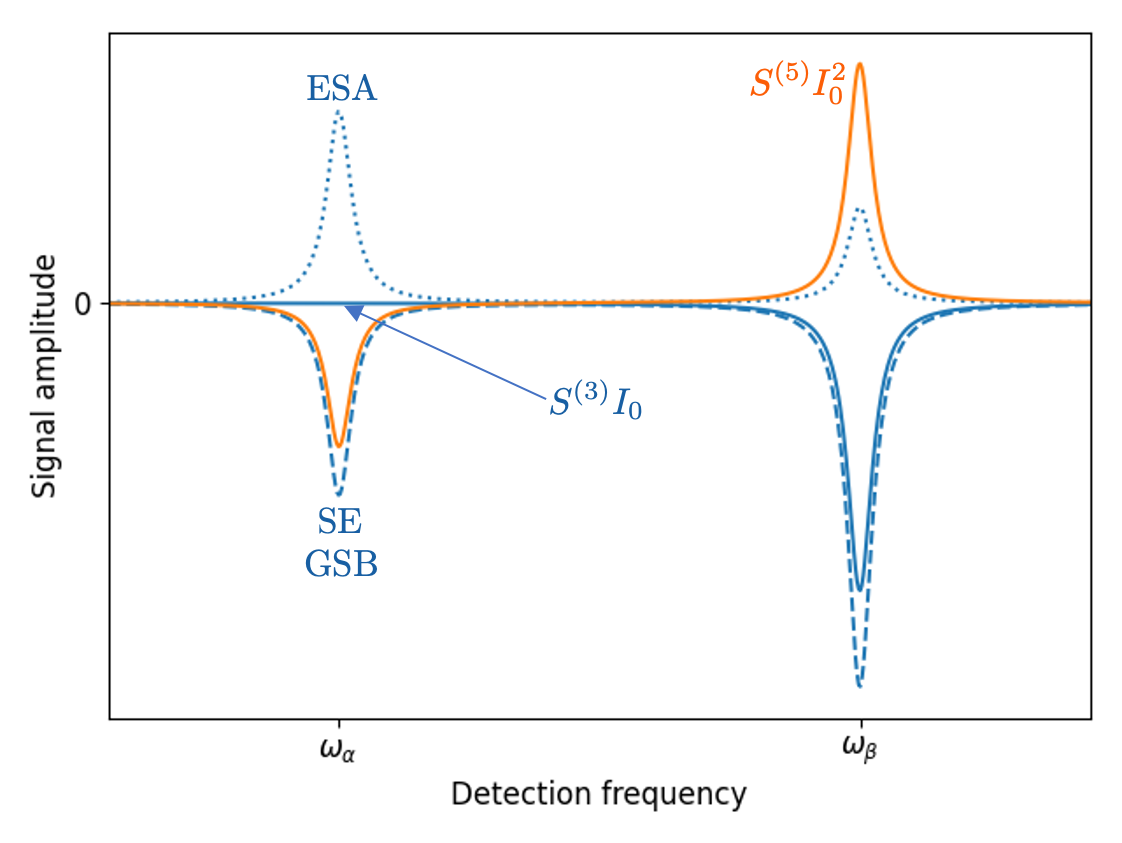}\caption{\label{fig:S3_and_S5}HOTA spectra for a model where the lower exciton
peak vanishes at third order (solid blue) and is negative at fifth
order (orange). The negative-signed $\text{GSB}+\text{SE}$ signals
(dashed) precisely cancel the positive-signed $\text{ESA}$ signal
(dotted). Here $J=0.18(\omega_{\beta}-\omega_{\alpha})$ and there
is no relaxation.}
\end{figure}

When the ratio of dipole couplings is correct, the third-order signal
at $\omega_{\alpha}$ can be found to cancel, as demonstrated in Fig.~\ref{fig:S3_and_S5}.
When $J=0.18\hbar(\omega_{B}-\omega_{A})$, the third-order signal
at $\omega_{\alpha}$ vanishes, while there is a negative-fifth order
signal. As in the ladder model, the new negative feature in the fifth
order cannot be an ESA$_{2}$ contribution but instead reports on
1-excitation signals that were masked in the third-order. The same
phenomenon can also occur in a model with $J=0$ but fast relaxation
from $\ket B$ to $\ket A$ \cite{maly2023}. In both realizations
of the biexciton model, one of the two exciton peaks is invisible
at third-order and appears as a negative peak in the fifth-order signal.
The simplicity of the two, quite different, parameter regimes under
which this phenomenon appears underscores the fact that this phenomenon
can occur in a wide range of systems.

This discussion emphasizes the value of measuring simultaneously both
$S^{(3)}$ and $S^{(5)}$, and shows some of the information that
can be revealed by comparing the two signals. We draw particular attention
to the fact that comparing the signs of peaks in third- and fifth-order
signals immediately reveals the nature of the dominant signal at each
spectral location.

Many methods have been used or proposed to extract different high-order
responses, frequently involving more than two pulses \cite{luttig2023a}.
Refs.~\citenum{yu2019,abramavicius2020} discuss 3-pulse experiments
that are high-order extensions of double-quantum (HODQ) spectroscopy
\cite{yang2008}. Both HOTA and HODQ extract the emission/absorption
lineshapes for higher lying excited states. HOTA also extracts population
dynamics for high-lying excited states. In contrast, HODQ does not
extract any population dynamics but rather extracts the frequency-frequency
correlation between excitation and emission at a population time of
0. HOTA is experimentally simpler because it requires only a standard
TA setup, whereas HODQ requires three pulses and phase-matching or
phase-cycling routines. HODQ has the advantage of being background-free,
as there is only a signal if higher-lying excited states exist\cite{abramavicius2020}.
In contrast HOTA always has non-vanishing signals $S^{(2n+1)}$ for
all $n$, even if the system only supports a single dipole-allowed
transition. There are also other methods for measuring higher-order
responses such as exciton-exciton interaction 2D spectroscopy (EEI2D),
a fifth-order signal that gives excitation and emission/absorption
lineshapes, as well as population dynamics, but is again more experimentally
challenging than HOTA \cite{maly2020}. Both HODQ and EEI2D potentially
suffer from higher-order contaminations, which can be difficult to
quantify. One of the key advantages offered by HOTA is the clean separation
of the nonlinear orders. 

We have laid the foundation for understanding the higher-order perturbative
terms that can now be measured using the method outlined in Ref.~\citenum{maly2023}.
We introduced a language to describe the processes contributing to
these higher-order signals and used two simple models to illustrate
the new kinds of spectral information that can be uncovered by comparing
the third- and fifth-order signals. In particular we showed that new
peaks emerging at fifth-order reveal new spectral information if the
peak is positive or reveal previously hidden spectral information
if the peak is negative.
\begin{acknowledgement}
We acknowledge helpful conversations with Tobias Brixner, Pavel Mal\'y,
and Julian L\"uttig and a careful reading by Julian L\"uttig. We
acknowledge funding from the Natural Sciences and Engineering Research
Council of Canada.
\end{acknowledgement}
\newpage
\renewcommand{\thefigure}{S\arabic{figure}}
\setcounter{figure}{0}
\renewcommand{\thetable}{S\arabic{table}}
\setcounter{table}{0}

\section*{Supplementary Information}

\subsection*{Pump-probe spectroscopy with THz pulses}

The expansion in Eq.~\ref{eq:perturbative-expansion} holds for any
wavelengths of the pump and probe beams, and we expect the decomposition
technique demonstrated in Ref.~\citenum{maly2023} to work regardless
of wavelength, from THz to xray. In the main text, we work in the
limit where the thermal energy is small compared to the laser center
frequency of both pump and probe, so that the initial system density
matrix exists exclusively in the $\ket 0$ states. At room temperature,
the thermal energy is equivalent to 6 THz. The use of THz probe pulses,
in conjunction with IR or optical pump pulses, is a widely used technique
\cite{george2008,xiao2022}, and the use of THz pump pulses is a growing
area of research \cite{hoffmann2009,lu2016,novelli2022}. We first
discuss the modifications required when using a THz probe and then
discuss the more significant changes when using a THz pump, as for
rotational spectroscopy. This discussion assumes ambient temperature,
but the relevant parameter is the ratio of the photon energy to the
thermal energy, and at other temperatures the THz scale should be
shifted accordingly.

\begin{figure}
\includegraphics{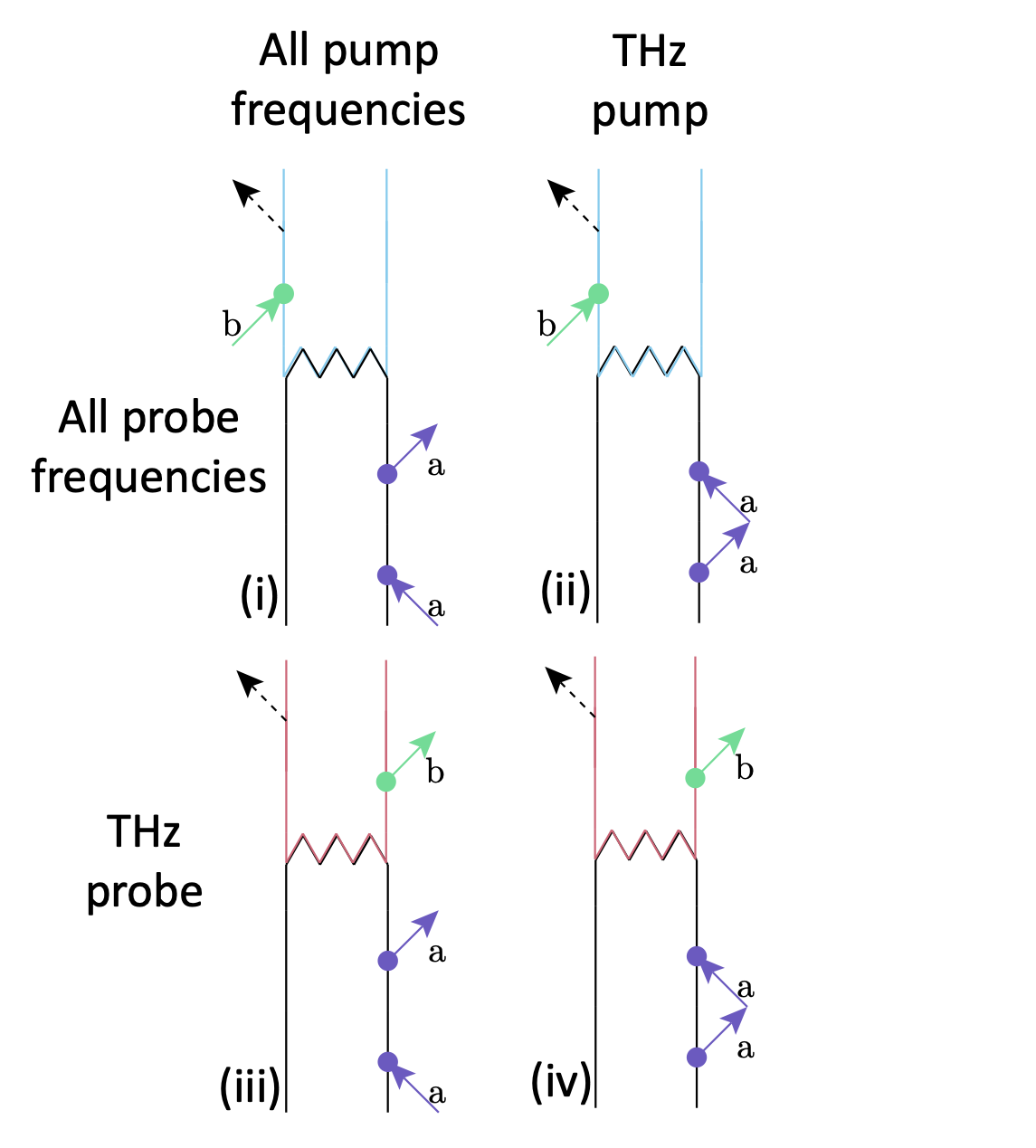}\caption{\label{fig:THz-diagrams} Example diagrams that contribute depending
upon the frequencies of the pump and probe pulses: (i) a ground-state
bleach diagram that always contributes to TA, (ii) a similar diagram
that only contributes when the probe includes THz frequencies, (iii)
a permutation of (i) that only contributes when the pump includes
THz frequencies, and (iv) a permutation of (iii) that only contributes
when both the pump and the probe include THz frequencies.}
\end{figure}

If the pump central frequency is well above thermal energy (e.g.,
mid-IR or shorter wavelengths), and the probe pulse is in the THz
regime, the interpretations and formalism introduced in the main paper
all still apply, but additional Feynman diagrams must be included,
which are formed by taking any of the diagrams in Figs.~\ref{fig:5th-order}
and \ref{fig:7th-order} that end in the states $\ket 0\bra 0$ and
replacing the absorption cap with the emission cap. These emission-type
diagrams occur because, even in the absence of a pump pulse, the probe
pulse can stimulate emission from thermally excited populations. Thus
the linear response of the system involves not only absorption but
also emission, which we call $\text{E}$. Thus there will also be
$\text{NE}^{(3)}$, $\text{E}^{(5)}$ and $\text{NE}^{(7)}$ processes
at third, fifth and seventh order, respectively, and so on. Figure~\ref{fig:THz-diagrams}(i)
shows a typical ground-state bleach ($\text{NA}^{(3)}$) diagram,
and Fig.~\ref{fig:THz-diagrams}(iii) shows the emission ($\text{NE}^{(3)}$)
counterpart of that diagram. For another example of an $\text{NE}^{(3)}$
Feynman diagram, see Fig.~17(viii) of Ref.~\citenum{lu2018}.

When the pump central frequency and thermal energy become comparable,
the interpretation of HOTA becomes more complicated since the excitation
number $n$ (shown schematically in the inset of Fig.~\ref{fig:SEnESAn})
cannot be well defined. These complications are similar to those in
standard TA interpretation. Depending upon the sensitivity of the
experiment, information from the lowest $N$ excitation states is
already present in the third-order signal. The initial state of the
system is a thermal average over multiple values of $n$. If we consider
a harmonic ladder with frequency $\omega$ where $\hbar\omega=k_{B}T_{sys}$,
where $k_{B}$ is Boltzmann's constant and $T_{sys}$ is the system
temperature, then the initial density matrix $\rho\propto e^{-H/k_{B}T}$
has $\ket 1$ states suppressed by a factor of $e$ and $\ket 2$
states suppressed by a factor of $e^{2}$. The standard third-order
TA signal then contains not only the dynamics and spectral features
of $\ket 1$ but also information from $\ket 2$ that is suppressed
by a factor of $e$ and information from $\ket 3$ that is suppressed
by $e^{2}$, and in general information from $\ket n$ that is suppressed
by a factor of $e^{n-1}$. If such a third-order experiment can resolve
signals up to $N$ excitations, then the fifth-order signal includes
information about the lowest $N+1$ excitation states. Thus we recover
an analogous argument to that given in the main text, where each higher
order signal adds one new excitation state.

This complication can be seen in the greatly increased number of diagrams
that must be considered for the case where $\hbar\omega_{\text{pump}}\lessapprox k_{B}T_{sys}$,
owing to the thermal average over multiple $\ket n$ in the initial
thermal state of the system; see Table~\ref{tab:THz-diagrams} for
the number of diagrams that contribute to various HOTA signals. Figure~\ref{fig:THz-diagrams}(ii)
shows an example of one additional diagram that must be considered,
if the pump pulse includes THz frequencies; it is a permutation of
a typical ground-state bleach diagram shown in Fig.~\ref{fig:THz-diagrams}(i),
but involves first de-excitation, followed by excitation, of the bra-side
of the density matrix. Figure~\ref{fig:THz-diagrams}(iv) shows an
additional diagram that must be considered if both the pump and probe
include THz frequencies. For an example of an analogous fifth-order
diagram that involves multiple de-excitations, see Fig.~17(xiv) of
Ref.~\citenum{lu2018}. The Diagram Generator described in Ref.~\citenum{rose2021a}
can be used to visualize all of the diagrams that contribute when
the pump and probe pulses both include THz frequencies and can also
be used to automate the resulting calculations of the spectra.  
\begin{table}
\begin{tabular}{ccc}
 & >THz pump & THz pump\tabularnewline
\hline 
\hline 
>THz probe & %
\begin{tabular}{c}
$\phantom{{0\}}}$\tabularnewline
6\tabularnewline
54\tabularnewline
540\tabularnewline
$\phantom{0}$\tabularnewline
\end{tabular} & %
\begin{tabular}{c}
$\phantom{0}$\tabularnewline
8\tabularnewline
96\tabularnewline
1280\tabularnewline
$\phantom{0}$\tabularnewline
\end{tabular}\tabularnewline
THz probe & %
\begin{tabular}{c}
8\tabularnewline
64\tabularnewline
640\tabularnewline
\end{tabular} & %
\begin{tabular}{c}
16\tabularnewline
192\tabularnewline
2560\tabularnewline
\end{tabular}\tabularnewline
\hline 
\end{tabular}\caption{\label{tab:THz-diagrams}Number of diagrams for calculating HOTA spectra
depending upon whether one or both of the pulses is at THz frequencies.
In each entry, the numbers given are for third, fifth, and seventh
orders, assuming the pump and probe pulses do not overlap in time.
Additional diagrams are required if the pulses overlap.}
\end{table}

While the diagrammatic details are different for THz than they are
for optical high-order TA spectroscopy, the main conclusions still
hold: higher-order signals give dynamical information about higher
excitation states, and each higher order similarly has access to new
spectral information.

\subsection*{Diagram Permutations}

\begin{figure}
\includegraphics{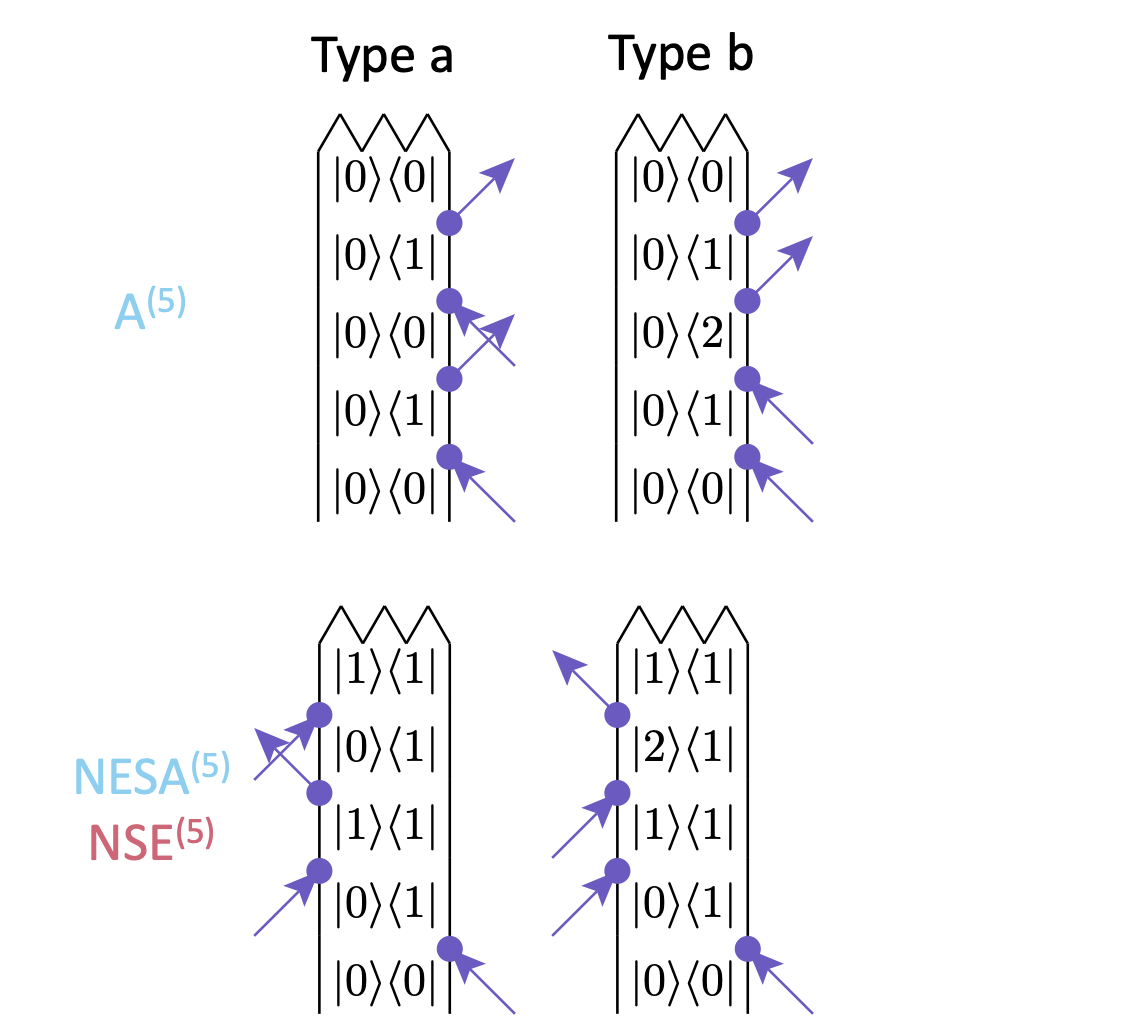}\caption{\label{fig:permutations}A permutation (left) of the interactions
in a base originally shown in Fig.~\ref{fig:5th-order} (right).
The top row shows two diagrams that contribute to the $\text{A}^{(5)}$
pathway, while the bottom row shows two diagrams that contribute to
the $\text{NESA}^{(5)}/\text{NSE}^{(5)}$ pathways.}
\end{figure}

Rather than display all of the diagrams, we have displayed representative
diagrams in Figures \ref{fig:SEnESAn}, \ref{fig:5th-order}, and
\ref{fig:7th-order}. All of the diagrams contributing to $S^{(3)}-S^{(7)}$
can be obtained from the displayed bases and caps by permuting of
the arrows of each base. As an illustration, Fig.~\ref{fig:permutations}
shows one permutation of the interactions in an $\text{A}^{(5)}$
diagram and one permutation in an $\text{NSE}^{(5)}$ diagram, alongside
the corresponding diagrams that were displayed in Fig.~\ref{fig:5th-order}
in the main text. Figure \ref{fig:permutations} shows permutation
of arrows occurring on the same side of the diagram, which show how
permuting the interactions on the same side of the diagram can change
which transition dipoles contribute to the diagram. In order to obtain
all diagrams, one must permute the order of the interactions on opposite
sides of the density matrix as well. Such permutations do not change
the diagram contributions in the impulsive limit but can change calculations
with finite-duration pulses, since different coherences are probed
during the pulse duration.

In what we call ``Type $a$'' diagrams, the system cycles between
only the $\ket 0$ and $\ket 1$ states, whereas in the ``Type $b$''
diagrams, the system transitions between the $\ket 0$, $\ket 1$,
and $\ket 2$ states. We note that this taxonomy, with the exception
of type $a$, requires a more complicated notation to extend to seventh
order and beyond. However, every order has diagrams that could be
considered type $a$, which involve only the dipole transition between
the $\ket 0$ and $\ket 1$ states. Such diagrams offer insight as
to why the HOTA signals are non-zero even for a two-level system,
where there are no higher-lying excitations present in the system.

The categorization of type $a$ and type $b$ is also useful in the
derivation of the biexciton signal strengths in Eqs.~\ref{eq:S^3}
and \ref{eq:S^5} in the main text. These equations are derived briefly
below.

\subsection*{Optical Biexciton Signal}

In Tables~\ref{tab:3rd-order} and \ref{tab:5th-order} we show the
spectral positions and weights for the third- and fifth-order signals,
respectively, for a generic ladder model. Adding up the signal contributions
from the column ``Total weight'' gives Eqs.~\ref{eq:S^3} and \ref{eq:S^5}.
As described just above, Table~\ref{tab:5th-order} distinguishes
between $a$ pathways, which involve only dipole transitions between
the $\ket 0$ and $\ket 1$ states during the pump pulse, and $b$
pathways, which involve dipole transitions between $\ket 0$, $\ket 1$,
and $\ket 2$ during the pump pulse, as shown in Fig.~\ref{fig:permutations}.

\begin{table}
\caption{\label{tab:3rd-order}Third-order process weights for the ladder model
of Eqs.~\ref{eq:H_ladder}, \ref{eq:mu_ladder}.}

\begin{tabular}{cccc}
 &  &  & \tabularnewline
\hline 
\hline 
Process & \# of & Total & Spectral\tabularnewline
 & diagrams & weight & position\tabularnewline
\hline 
$\text{GSB}$ & 2 & $-\mu_{10}^{4}$ & $\omega_{10}$\tabularnewline
$\text{SE}$ & 2 & $-\mu_{10}^{4}$ & $\omega_{10}$\tabularnewline
$\text{ESA}$ & 2 & $+\mu_{21}^{2}\mu_{10}^{2}$ & $\omega_{21}$\tabularnewline
\hline 
 &  &  & \tabularnewline
\end{tabular}
\end{table}

\begin{table}
\caption{\label{tab:5th-order}Fifth-order diagram weights for the ladder model
of Eqs.~\ref{eq:H_ladder}, \ref{eq:mu_ladder}.}

\begin{tabular}{cccc}
 &  &  & \tabularnewline
\hline 
\hline 
Process & \# of & Total & Spectral\tabularnewline
 & diagrams & weight & position\tabularnewline
\hline 
$\text{A}_{a}^{(5)}$ & 8 & $+\mu_{10}^{6}/3$ & $\omega_{10}$\tabularnewline
$\text{A}_{b}^{(5)}$ & 2 & $+\mu_{21}^{2}\mu_{10}^{4}/12$ & $\omega_{10}$\tabularnewline
$\text{NSE}_{a}$ & 8 & $+\mu_{10}^{6}/3$ & $\omega_{10}$\tabularnewline
$\text{NSE}_{b}$ & 8 & $+\mu_{21}^{2}\mu_{10}^{4}/3$ & $\omega_{10}$\tabularnewline
$\text{NESA}_{a}$ & 8 & $-\mu_{21}^{2}\mu_{10}^{4}/3$ & $\omega_{21}$\tabularnewline
$\text{NESA}_{b}$ & 8 & $-\mu_{21}^{4}\mu_{10}^{2}/3$ & $\omega_{21}$\tabularnewline
$\text{SE}_{2}$ & 6 & $-\mu_{21}^{4}\mu_{10}^{2}/4$ & $\omega_{21}$\tabularnewline
$\text{ESA}_{2}$ & 6 & $+\mu_{32}^{2}\mu_{21}^{2}\mu_{10}^{2}/4$ & $\omega_{32}$\tabularnewline
\hline 
 &  &  & \tabularnewline
\end{tabular}
\end{table}

\bibliography{HigherOrderInterpretation}

\end{document}